\documentstyle[aps,prl,psfig,twocolumn]{revtex}
\begin{document}
\draft
\title{Renormalization Group Approach to Non--equilibrium Green 
Functions in Correlated Impurity Systems}
\author
{T. A. Costi\cite{tac-email}}
\address
{Institut Laue-Langevin, B.P.156 38042 Grenoble, 
Cedex 9, France.\\
Universit\"{a}t Karlsruhe, Institut f\"{u}r Theorie der Kondensierten
Materie, 76128 Karlsruhe, Germany.}
\maketitle
\begin{abstract}
We present a new technique for calculating non--equilibrium Green
functions for impurity systems with local interactions. We use an analogy to
the calculation of response functions in the x--ray problem.
The initial state and the final state problems, which correspond to the
situations before and after the disturbance (an electric or magnetic
field, for example) is suddenly switched on, are solved with the aid of
Wilson's momentum shell renormalization group. The method
is illustrated by calculating the non--equilibrium dynamics of the
ohmic two--state problem.
\end{abstract}
\section{Introduction}
Recently there has been interest in the non--equilibrium
transport properties of small devices, such as quantum dots and
ultrasmall tunnel junctions \cite{qds.95,meso.91}. These systems, 
together with some resonant
tunneling devices \cite{eaves.94}, 
offer new possibilities for studying many--body effects
due to strong local Coulomb interactions. The importance of these interactions
in small devices is seen, for example, in the suppression of tunneling or
Coulomb Blockade effect \cite{girvin.90}.
Phenomena such as the Kondo effect in quantum dots and the Fermi edge 
singularity in resonant tunneling devices have been predicted 
\cite{matveev.92,ng.88,glazmann.88,meir.91+93,wingreen.94,hershfield.91+92,schoeller.96} 
and some aspects of these have been confirmed experimentally \cite{eaves.94}. 
The usual starting point for dealing
with non--equilibrium transport in such 
systems has been the formalism developed by Keldysh
\cite{keldysh.65} and Kadanoff and Baym \cite{kadanoff.62}. 
Below we describe a non--perturbative approach based on the 
numerical renormalization group (NRG) method \cite{wilson.75,kww.80}
which allows the calculation of non--equilibrium Green functions for 
the above systems. We consider only the case in which the perturbation
(an electric or magnetic field)
causing the non--equilibrium effects is suddenly switched on at time $t=0$.
The non--equilibrium Green Functions will then be calculated by 
solving an initial ($t<0$) and final ($t>0$) state problem as in 
the case of calculating 
the photoemission and absorption spectra in the x--ray problem \cite{nd.69}. 
In this paper we concentrate on calculating the non--equilibrium 
properties of 
a specific model, the Ohmic two state system \cite{caldeira+leggett}. The 
application of the method to the systems mentioned 
above follows along the same lines, the only difference being the solution, 
using the NRG method, of different initial and final state Hamiltonians.

The paper is organized as follows: in Sec. II we introduce the standard 
model of the Ohmic two state system, formulate the problem of calculating 
the non--equilibrium dynamics of this model in terms of solving an initial
and a final state problem and introduce an equivalent model, the anisotropic
Kondo model, which we actually use in the calculations. In Sec. III we 
describe the NRG, its application to dynamical quantities and an 
approximate evaluation of the formally exact expressions for 
the non--equilibrium quantities. An exact evaluation of 
non--equilibrium quantities first has to overcome certain technical 
difficulties, which we describe, and is postponed for the future.
Sec. IV contains our results for the non--equilibrium dynamics of the
Ohmic two--state system, obtained on the basis of NRG calculations for
the anisotropic Kondo model. In Sec. V we summarize our main conclusions.

\section{Formulation}
\subsection{The Ohmic Two--State System}
The non--equilibrium properties of the Ohmic two state problem 
are of main interest in macroscopic quantum 
coherence experiments in rf SQUIDs \cite{chakravarty.83}. 
Typically, an rf SQUID can be in one of two possible fluxoid states. 
By applying a bias (corresponding to an external magnetic field), 
for times $t<0$, the system is prepared in one of the two states.
The dynamics after the bias is removed at $t>0$ is then intrinsically
a non--equilibrium property. The Hamiltonian, $H$, of the system is 
time dependent with a sudden perturbation at $t=0$, so that we can
write $H(t)=[1-\theta(t)]H_{I}+\theta(t)H_{F}$ 
where $H_{I}, H_{F}$ are the Hamiltonians before and after the bias 
is switched off. The Hamiltonian $H_{I,F}$ describing the Ohmic two
state system is given by the spin--boson model \cite{leggett.87},

\begin{eqnarray}
H_{SB} & = &-\frac{1}{2}\hbar\Delta \sigma_{x}+\frac{1}{2}\epsilon\sigma_{z}
        +\sum_{\alpha} \omega_{\alpha}(a_{\alpha}^{\dagger}a_{\alpha}+\frac{1}{2})\nonumber\\
        &+&\frac{1}{2}q_{0}\sigma_{z}\sum_{\alpha}
\frac{C_{\alpha}}{\sqrt{2m_{\alpha}\omega_{\alpha}}}(a_{\alpha}+a_{\alpha}^{\dagger})\label{eq:SB}.
\end{eqnarray}
Here $\sigma_{i},i=x,y,z$ are Pauli spin matrices, the two states of the
system correspond to $\sigma_{z}=\uparrow$ and $\sigma_{z}=\downarrow$ 
(i.e. $\sigma_{z}=\uparrow,\downarrow$ correspond to the two possible
fluxoid states of the rf SQUID). 
$\Delta$ is the bare tunneling matrix element and $\epsilon$ is a bias.
The environment is represented by an infinite set of 
harmonic oscillators (labeled by the index $\alpha$) 
with masses $m_{\alpha}$ and 
frequency spectrum $\omega_{\alpha}$ coupling
linearly to the coordinate $Q=\frac{1}{2}q_{0}\sigma_{z}$ 
of the two--level system via a
term characterized by the couplings $C_{\alpha}$
(the two--level system coordinate could be the total flux, 
$\phi=\phi_{1}\sigma_z$, in the case of an rf SQUID experiment). 
The environment spectral
function is given in terms of these couplings, oscillator masses 
and frequencies by
 $J(\omega)=\frac{\pi}{2}
\sum_{\alpha}(\frac{C_{\alpha}^{2}}{m_{\alpha}\omega_{\alpha}})
\delta(\omega-\omega_{\alpha})$.
In the case of an Ohmic heat bath, of interest to us here, we have 
$J(\omega)=2\pi\alpha\omega$, for $\omega << \omega_{c}$, where $\omega_{c}$
is a high energy cut--off and $\alpha$ is a dimensionless parameter 
characterizing the strength of the dissipation. This form for the spectral
function is appropriate for
describing quantum dissipation in an rf SQUID. Preparation of the
system in a state with $\sigma_{z}=+1$ with the oscillators relaxed about
this state is equivalent to setting $\epsilon=-\infty$ in $H_{SB}$, so the
initial state problem corresponds to solving $H_{I}=H_{SB}(\epsilon=-\infty)$.
Similarly, the final state problem is given by $H_{F}=H_{SB}(\epsilon=0)$.
The Ohmic spin--boson model has been intensively studied (for reviews we 
refer the reader to \cite{leggett.87,weiss.93}). We outline some 
of its features in order to introduce some useful notation. The model has
a low energy scale, $\Delta_r<\Delta$ for $\Delta << \omega_c$, 
which depends on the dissipation
strength $\alpha$, and which may be interpreted as a renormalized tunneling
amplitude. For $\alpha<< 1$ the dynamics corresponds to damped oscillations,
with a crossover to incoherent behaviour with increasing dissipation strength.
For $\alpha\rightarrow 1^{-}$, the renormalized tunneling amplitude vanishes
giving rise to the phenomenon of ``localization'' or ``self--trapping'' for 
$\alpha > \alpha_{c}\approx 1$ ($\alpha_{c}$ depends also on the precise 
value of $\Delta$). The dynamical quantities exhibiting the above features
are defined below.

\subsection{Non--equilibrium Dynamical Quantities}
The simplest non--equilibrium dynamical quantity to study for the spin--boson
model is the quantity $P(t)=\langle \sigma_{z}(t)\rangle_{\rho_I}$ 
\cite{leggett.87} where the
thermodynamic average is taken with respect to the initial density matrix
$\rho_I(\beta)=e^{-\beta H_{I}}/Tr{\;e^{-\beta H_{I}}}$, 
$\beta$ is the inverse temperature, and
the time evolution is with respect to the Hamiltonian after the bias is
switched off at time $t=0$, i.e. $\sigma_{z}(t)=e^{iH_{F}t}\sigma_{z} 
e^{-iH_{F}t}$. Hence, $P(t) = 1$ for $t<0$ due to 
the infinite bias $\epsilon=-\infty$, and for $t>0$,
when the bias is switched off ($\epsilon=0$), $P(t)$ describes how 
the two--level system
co--ordinate $\sigma_z$ relaxes
to its long--time value of zero.
Another quantity of interest is the retarded two time 
Green function, 
$G_{r}(t,t')=-i\theta(t-t')\langle[\sigma_{z}(t),\sigma_{z}(t')]
\rangle_{\rho_{I}}$, with the thermodynamic average defined as above. 
Since time translational invariance is broken, $G_{r}(t,t')$ depends on 
both times explicitly. We consider the Fourier transform of
$G_{r}(t,t')$ with respect to both the sum $t+t'$ and difference $t-t'$ of
the time variables. The resulting spectral density 
$C_{r}(\omega,\Omega)=-\frac{1}{\pi}\mbox{Im}\;G_{r}(\omega+i\delta,\Omega)$,
with $\omega,\Omega$ the Fourier frequency variables corresponding to 
$t-t',t+t'$, is given within a Lehmann representation by the following
expression,

\begin{eqnarray}
C_{r}(\omega,\Omega) & = & 
\frac{2\pi}{Z_{I}}\sum_{m_{I},m_{F},m_{F}',m_{F}''} e^{-\beta E_{m_{I}}}
\langle m_{I}|m_{F} \rangle\langle m_{F}''|m_{I}\rangle\nonumber\\
                     & \times &
\langle m_{F}|\sigma_{z}|m_{F}'\rangle
\langle m_{F}'|\sigma_{z}|m_{F}''\rangle 
\delta(\Omega-\frac{E_{m_F}-E_{m_{F}''}}{2})\nonumber\\
& \times & 
[\delta(\omega+[\frac{E_{m_{F}}+E_{m_{F}''}}{2}-E_{m_{F}'}])\nonumber\\
& - & \delta(\omega-[\frac{E_{m_{F}}+E_{m_{F}''}}{2}-E_{m_{F}'}])].
\label{eq:cneq}
\end{eqnarray}
Here, $E_{m_{I}}$ and $|m_{I}\rangle$ are the many--body 
eigenvalues and eigenstates
of the initial state Hamiltonian $H_{I}$, $Z_{I}$ the corresponding partition
function and $E_{m_{F}}$, $|m_{F}\rangle$,
$E_{m_{F}'}$, $|m_{F}'\rangle,\dots $ 
the many--body eigenvalues and eigenstates
of the final state Hamiltonian $H_{F}$. In the equilibrium case, 
$H_{I}=H_{F}=H$, and the corresponding spectral density 
$C_{r}^{eq}(\omega,\Omega)$ reduces to,

\begin{eqnarray}
\lefteqn{C_{r}^{eq}(\omega,\Omega)  =  
\frac{2\pi}{Z}\sum_{m,m'} e^{-\beta E_{m}}
|\langle m|\sigma_{z}|m'\rangle|^{2}\delta(\Omega)}\nonumber\\
&\times& 
[\delta(\omega+[E_{m}-E_{m'}]) - \delta(\omega-[E_{m}-E_{m'}])],\label{eq:ceq}
\end{eqnarray}
where $E_{m}, |m\rangle$ are the many--body eigenvalues and eigenfunctions
of $H$ and $Z$ is the corresponding partition function (the delta function,
$\delta(\Omega)$, in the above expression reflects the fact that in the
equilibrium case, $G_{r}(t,t')$ depends only on the difference of the 
time variables).

We see that the non--equilibrium
spectral density differs from the equilibrium one in several ways: first,
even at $T=0$, no groundstate energy appears in the delta functions, the
excitations are between arbitrary (final) excited states of the system. 
This reflects the fact that there is no stationary groundstate 
for a non--equilibrium 
situation. Secondly, the non--equilibrium aspects, which are a result of
an initial state preparation, are reflected in the presence of overlap
matrix elements between the initial and final states. Finally the dependence
on $\Omega$ is a measure of the importance of transient effects. Neglecting
these effects results in the following simplified expression for the
spectral density,
\begin{eqnarray}
\lefteqn{C_{0r}(\omega)  =  C_{r}(\omega,\Omega=0)}\nonumber\\ 
& = &\frac{2\pi}{Z_{I}}\sum_{m_{I},m_{F},m_{F}'} e^{-\beta E_{m_{I}}}
|\langle m_{I}|m_{F} \rangle|^{2}
|\langle m_{F}|\sigma_{z}|m_{F}'\rangle|^{2}\delta(\Omega)\nonumber\\
& \times & 
[\delta(\omega+[E_{m_{F}}-E_{m_{F}'}]) - 
\delta(\omega-[E_{m_{F}}-E_{m_{F}'}])].\label{eq:c0r}
\end{eqnarray}
This describes the steady state case $t+t'\rightarrow\infty$. In a strict
sense it is not a non--equilibrium quantity, although it does take into
account the effects of an initial state preparation on the 
correlation function $\langle\sigma_{z}(t)\sigma_{z}(0)\rangle$. Our 
motivation for calculating this quantity is simply to illustrate that
our technique applies also to two--time Green functions. The calculation
of the full spectral density $C_{r}(\omega, \Omega)$, including both 
frequencies involves a straightforward generalization.

Similarly we can write the Fourier transform of $P(t)$, within a Lehmann
representation, as,
\begin{eqnarray}
P(\omega) & = & \frac{1}{Z_{I}}\sum_{m_{I},m_{F},m_{F}'}e^{-\beta E_{m_{I}}}
\langle m_{I}|m_{F} \rangle \langle m_{F}'|m_{I} \rangle\nonumber\\
& \times &
\langle m_{F}|\sigma_{z}|m_{F}'\rangle\delta(\omega-(E_{m_{F}}-E_{m_{F}'})),
\label{eq:pw}
\end{eqnarray}
where the same notation as above is used.

$P(t)=\int_{0}^{\infty}P(\omega)\cos(\omega t)dt$ 
contains information on the onset
of quantum oscillations in the two level system. For small values
of the dissipation strength, $\alpha << 1$, it is known  from
the ``Non--Interacting Blip Approximation'' (NIBA)\cite{leggett.87} 
that $P(t)$ exhibits damped oscillations with a renormalized tunneling
frequency 
$\Delta_{r}=\Delta[\frac{\Delta}{\omega_{c}}]^{\frac{\alpha}{1-\alpha}}$. 
$P(\omega)$ will exhibit two peaks at $\omega=\pm \Delta_{r}$. At the
exactly solvable Toulouse point\cite{toulouse.69}, 
$\alpha=1/2$, where the Ohmic two--state
system reduces to the resonant level model \cite{leggett.87,vigmann.78}, 
the dynamics is incoherent and
$P(t)$ decays exponentially. $P(\omega)$ consists of a single peak
at $\omega=0$. It is not clear at which value of the
dissipation strength the crossover to incoherent behaviour occurs, in
particular whether it occurs at exactly $\alpha=1/2$ or for some smaller
value of $\alpha$. This may depend on the definition of the crossover 
and on whether equilibrium or non--equilibrium quantities are being
studied. For equilibrium quantities a smooth 
crossover has been found to occur at $\alpha=1/3$ \cite{costi.96,lesage.96}. 

\subsection{The Anisotropic Kondo Model}

Instead of solving directly the spin--boson model with the NRG method it
is more convenient to solve an equivalent fermionic model which has the same
dynamics. This is the Anisotropic Kondo Model (AKM). The equivalence
has been shown at the Hamiltonian level via bosonization \cite{guinea.85b}. 
This was believed to be valid in the region $\alpha > 1/2$, which 
corresponds (see below for the precise statement of the equivalence) 
to the region in the parameter space of the AKM between weak--coupling 
and the Toulouse point. In fact, recent work \cite{costi.96} shows 
that the equivalence extends beyond the Toulouse point into the region 
describing weak dissipation $0<\alpha<1/2$ 
(or large antiferromagnetic $J_{\parallel}$ in the AKM, see also 
\cite{kotliar.96}). 
The AKM is given by \cite{anderson.70}
\begin{eqnarray}
H &=& \sum_{k,\sigma} \epsilon_{k}c_{k\sigma}^{\dagger}c_{k\sigma} + 
\frac{J_{\perp}}{2}\sum_{kk'}
        (c_{k\uparrow}^{\dagger}c_{k'\downarrow}S^{-} +
         c_{k\downarrow}^{\dagger}c_{k'\uparrow}S^{+})\nonumber\\
  &+& \frac{J_{\parallel}}{2}\sum_{kk'}
         (c_{k\uparrow}^{\dagger}c_{k'\uparrow} -
          c_{k\downarrow}^{\dagger}c_{k'\downarrow})S^{z} 
+ g\mu_{B}hS_{z}.\label{eq:AKM}
\end{eqnarray}
The first term represents non--interacting conduction electrons and the
second and third terms represent an exchange interaction between a localized
spin $1/2$ and the conduction electrons with strength 
$J_{\perp},J_{\parallel}$. 
A local magnetic field, $h$, coupling only 
to the impurity spin in the Kondo model (the last term in Eq.~\ref{eq:AKM})  
corresponds to a finite bias, $\epsilon$, in 
the spin--boson model. The correspondence 
between $H$ and $H_{SB}$ is then given by $\epsilon=g\mu_{B}h$,  
$\frac{\Delta}{\omega_{c}}= \rho J_{\perp}$ and 
$\alpha=(1+ \frac{2 \delta}{ \pi})^{2}$, where
$\tan{\delta}= -\frac{ \pi \rho J_{\parallel}}{4}$, $\delta$ is the phase 
shift for scattering of electrons from a potential $J_{\parallel}/4$ and 
$\rho=1/2D$ is the conduction electron density of states per
spin at the Fermi level for a flat band of width $2D$ 
\cite{leggett.87,costi.96}. 
We note that weak dissipation ($\alpha\rightarrow 0$) in the spin--boson
model corresponds to extreme anisotropy ($J_{\parallel}\rightarrow\infty$)
in the Kondo model. For zero dissipation, $J_{\parallel}=\infty$, the two
states $\psi_{\pm}=\frac{1}{\sqrt{2}}(|\uparrow\rangle|\downarrow\rangle_{0}
\pm|\downarrow\rangle|\uparrow\rangle_{0})$ made up from the impurity states
and the local conduction electron Wannier orbitals 
$|\sigma\rangle_{0} =\sum_{k}c_{k\sigma}^{\dagger}|vac\rangle$, where 
$|vac\rangle$ is the vacuum, are split by $J_{\perp}=\Delta$ (with the
identification $\omega_{c}=2D$) and are completely decoupled from the rest
of the conduction band, thus forming an isolated two--level system. The 
system exhibits coherent oscillations with $P(t)=\cos(\Delta t)$. As 
$J_{\parallel}$ is decreased from $+\infty$, the two--levels become weakly
coupled, with strength $\frac{D^{2}}{J_{\parallel}}\propto \alpha$, 
for $\rho J_{\parallel}>>1$, to the remaining conduction 
states and their splitting is renormalized downwards.
The low energy scale of the model is given by the Kondo temperature for the
anisotropic Kondo model, \cite{tsvellick.83},
$T_{K}(J_{\perp},J_{\parallel})<J_{\perp}$, for $J_{\perp}<<D$, which in the
language of the dissipative two--state system corresponds to a renormalized
tunneling amplitude $\Delta_{r}$ \cite{costi.96}. 
An extensive discussion of the 
equivalence between the anisotropic Kondo model and the Ohmic two--state
system is given elsewhere.

\section{Calculation of Non--equilibrium Dynamics via  the NRG}
\subsection{The NRG}
The dynamical quantities $P(t)$ and $G_{r}(t,t')$ defined
above for the two--level system translate into the corresponding quantities
for the Kondo model ($\sigma_{z}\rightarrow S_{z}$ under the equivalence).
We calculate these quantities by applying Wilson's momentum shell 
renormalization group method generalized to the calculation
of dynamical quantities (e.g. \cite{costi.96,costi.94}). 
Thus in addition to
solving an initial state problem $H_{I}=H_{AKM}(\epsilon=-\infty)$ and
a final state problem $H_{F}=H_{AKM}(\epsilon=0)$, the final state 
matrix elements of the variable $\sigma_{z}$ and the overlap matrix
elements appearing in the above expressions for 
$P(\omega),C_{r}(\omega)$ are also calculated. The diagonalization
of $H_{AKM}$ proceeds as follows (for full details see \cite{wilson.75}),
(i) the spectrum is linearized about the Fermi energy 
$\epsilon_{k}\rightarrow v_{F}k$, 
(ii) a logarithmic mesh of $k$ points $k_{n}=\Lambda^{-n}$ is introduced
to achieve a separation of energy scales, and (iii) a unitary 
transformation of the $c_{k\sigma}$ is made such
that $f_{0\sigma}=\sum_{k}c_{k\sigma}$ is the first operator in a new 
basis, $f_{n\sigma},n=0,1,\dots$, which tridiagonalizes 
$H_{c}=\sum_{k\mu}\epsilon_{k\mu}c_{k\mu}^{\dagger}c_{k\mu}$ 
in k--space, i.e. 
$H_{c}\rightarrow \sum_{\mu}\sum_{n=0}^{\infty}\xi_{n}\Lambda^{-n/2}
(f_{n+1\mu}^{\dagger}f_{n\mu}+ h.c.),$
with $\xi_{n}\rightarrow (1+\Lambda^{-1})/2$ for $n>>1$. 
The Hamiltonian (\ref{eq:AKM}) with the above discretized 
form of the kinetic energy is now diagonalized by the following iterative 
process: (a) one defines a sequence of finite size Hamiltonians 
$H_{N} = \sum_{\mu}\sum_{n=0}^{N-1}\xi_{n}\Lambda^{-n/2}
(f_{n+1\mu}^{\dagger}f_{n\mu}+ h.c.) + 
\frac{J_{\perp}}{2}
        (f_{0\uparrow}^{\dagger}f_{0\downarrow}S^{-} +
         f_{0\downarrow}^{\dagger}f_{0\uparrow}S^{+})
  + \frac{J_{\parallel}}{2}
         (f_{0\uparrow}^{\dagger}f_{0\uparrow} -
          f_{0\downarrow}^{\dagger}f_{0\downarrow})S^{z}$ for $N\ge 0$; 
(b) the Hamiltonians
$H_{N}$ are rescaled by $\Lambda^{\frac{N-1}{2}}$ such that the 
energy spacing remains the same, i.e. $\bar{H}_{N}=\Lambda^{\frac{N-1}{2}}
H_{N}$. This defines a renormalization group transformation
$\bar{H}_{N+1} = \Lambda^{1/2}\bar{H}_{N} + \sum_{\mu}\xi_{N}
(f_{N+1\mu}^{\dagger}f_{N\mu}+h.c.) - E_{G, N+1}$,
with $E_{G,N+1}$ chosen so that the ground state energy
of $\bar{H}_{N+1}$ is zero. Using this recurrence relation, the sequence
of Hamiltonians $\bar{H}_{N}$ for $N=0,1,\ldots$ is iteratively diagonalized.
This gives the excitations and many body eigenstates at a corresponding
set of energy scales $\omega_{N}$ defined by 
$\omega_{N}=\Lambda^{-\frac{N-1}{2}}$ and allows a direct calculation of
the dynamical quantities $P(t), G_{r}(t,t')$, or more precisely their Fourier
transforms using the Lehmann representations (\ref{eq:cneq},\ref{eq:pw}). 
Our results were obtained for $\Lambda=2$, keeping the 320 lowest states 
at each iteration. Truncating the spectrum in this way restricts the range
of excitations $\omega$ at iteration $N$ to be such that 
$\omega_{N}\leq \omega \leq K\omega_{N}$, where $K=K(\Lambda)\approx 7$
for $\Lambda=2$. In this paper we discuss only the $T=0$ results.

In diagonalizing the Hamiltonians
$\bar{H}_{N}$ the following symmetries are used to reduce the size of the
matrices, (i) conservation of z--component of total spin 
$S_{z}^{tot}=S_{z}+\sum_{n=0}^{N}\frac{1}{2}(
f_{n\uparrow}^{\dagger}f_{\uparrow}-f_{n\downarrow}^{\dagger}f_{\downarrow})$,
(ii) conservation of total pseudo--spin\cite{jones.87}, 
where the pseudo--spin operators $j^{+},j^{-},j_{z}$ are defined by 
$j^{+}=\sum_{n=0}^{N}(-1)^{n}f_{n\uparrow}^{\dagger}
f_{n\downarrow}^{\dagger}$, $j^{-}=(j^{+})^{\dagger}$ and 
$j_{z}=\sum_{n=0}^{N}(f_{n\mu}^{\dagger}f_{n\mu}-\frac{1}{2})$. These 
symmetries hold for both the zero and finite bias cases, which we consider
in this paper ($\epsilon=0$ for the final state problem and 
$\epsilon=-\infty$ for the initial state problem). 
At this point we mention that although it is in principal 
possible to apply this renormalization group method directly to the
spin--boson model, in practice there are disadvantages in the case of
the bosonic problem \cite{note1}.

\subsection{Broadening procedure for the discrete spectra}
We note that the use of a discretized model implies that at 
each iteration $P(\omega)$ is a series of delta functions. 
Smooth curves are obtained by broadening these delta functions with a
Gaussian of width appropriate to the level spacing of $H_{N}$\cite{costi.94}. 
The width of the Gaussians will not influence the intrinsic peak widths at
low energies, where the logarithmic spacing ensures high resolution, but
may do so at higher energies where the resolution is lower. This 
problem has been discussed in detail in 
\cite{yoshida.90}, where a refined NRG scheme to overcome it has been 
suggested. In most problems one is interested in the low energy 
behaviour and such a refinement is not required, however for the spin--boson
model considered here there is interesting dynamics at high energies and
such a refinement is required in order to obtain a complete description
of all aspects of the high energy (short time) dynamics (by high energy, we
mean high relative to the renormalized tunneling amplitude, but still low
relative to the high energy cut--off in the model). In the present 
paper we have not implemented the above refined scheme, so we shall state 
below which aspects of the high energy dynamics we are able to obtain within
the unmodified NRG scheme.

\subsection{Approximate evaluation of non--equilibrium quantities}
At this stage we point out an importance difference in the
calculation of non--equilibrium quantities with respect to equilibrium
quantities such as $C_{r}^{eq}(\omega,\Omega)$. At $T=0$, the latter
can be calculated on each energy scale $\omega_N$ by restricting attention
to a single energy shell $N$. This is due to the existence of a 
stationary groundstate, from which all excitations in the expression
for $C_{r}^{eq}$ can be measured. The delta functions in the
Lehmann representation, (\ref{eq:ceq}), then imply that in order to calculate 
$C_{r}^{eq}$ at frequency $\omega$, only one energy shell, 
that for which $\omega\approx\omega_{N}$, is required, i.e.
\begin{eqnarray}
\lefteqn{C_{r}^{eq}(\omega,\Omega,T=0)  =  
\frac{2\pi}{Z_{N}}\sum_{m'}
|\langle GS|\sigma_{z}|m'{\rangle}_{N}|^{2}\delta(\Omega)}\nonumber\\
& \times & 
[\delta(\omega+[E_{GS}^{N}-E_{m'}^{N}]) - 
\delta(\omega-[E_{GS}^{N}-E_{m'}^{N}])],
\label{eq:ceq-nrg}
\end{eqnarray}
where $|m'\rangle_{N}$, $|GS>_{N}$ are the eigenstates and the groundstate, 
corresponding to iteration $N$ in the NRG procedure, $E_{m'}^{N}$ and 
$E_{GS}^{N}$ the corresponding eigenvalues, and $Z_{N}$ the ($T=0$) 
partition function. Contributions from energy shells $n=0,1,\dots,N-1$
have $E_{m}^{n}-E_{GS}^{n}>\omega$ so they make no contribution to the 
spectral density $C_{r}^{eq}$ at frequency $\omega\approx\omega_N$. 
At finite temperatures, there can be contributions from higher energy
states, although the Boltzmann factors in (\ref{eq:ceq}) will suppress these.

The situation is different for non--equilibrium dynamical
quantities, such as $P(\omega)$. From the Lehmann representation (\ref{eq:pw})
we see that, even at $T=0$, no groundstate energy appears
in the delta functions. Instead the ($T=0$) excitations are between arbitrary
(final) states. The response at frequency $\omega\approx\omega_{N}$ can have
contributions from all energy shells $n=0,1,\dots,N$. Hence, 
if we evaluate $P(\omega)$, at the
frequency $\omega\sim \omega_{N}=D\Lambda^{-\frac{N-1}{2}}$, by taking
into account only the $N$'th energy shell, we have only the 
{\em approximate} result 
\begin{eqnarray}
\lefteqn{P(\omega)\approx 
P_{N}(\omega)}\nonumber\\
& = & \frac{1}{Z_{N,I}}\sum_{m_{F},m_{F}'}
\langle m_{I,GS}|m_{F} \rangle_{N} \langle m_{F}'|m_{I,GS} 
\rangle_{N}\nonumber\\ & \times &
\langle m_{F}|\sigma_{z}|m_{F}'\rangle_{N}
\delta(\omega-(E^{N}_{m_{F}}-E^{N}_{m_{F}'})).\label{eq:pw-approx}
\end{eqnarray}
The approximation implicit in this procedure is that small excitations
$\omega\sim E^{n}_{m_{F}}-E^{n}_{m_{F}'}$ between higher energy excited
states (i.e. from energy shells with $n<N$), make a negligible 
contribution to $P(\omega)$ compared to those on a scale 
$\omega\sim\omega_N$ from the $N$'th energy shell. This is clearly only
an approximation to the formally exact expression (\ref{eq:pw}). The
results and arguments to be presented below show that the above 
approximation has a regime of validity and a regime where it breaks down. 
We stress, however, that a full multiple--shell evaluation of (\ref{eq:pw}) 
will ultimately be required. This is currently not feasible within the standard
NRG procedure described in Sec.~III~A--B. One first has to overcome
two problems: (i) a possible double counting of 
excitations in adding contributions to $P(\omega)$ from 
different energy shells, 
(ii) a meaningful way of adding contributions to $P(\omega)$ from 
higher energy shells which have different resolutions.
The second problem can be overcome by
replacing the standard NRG procedure of Sec.~III~A--B by one which
eliminates any dependence of static and dynamic quantities on the 
logarithmic discretization \cite{yoshida.90}.
These problems do not arise in the case of equilibrium dynamical
quantities since, as discussed above, the spectral densities can be
calculated essentially without approximation by restricting attention to
just one energy shell for each frequency of interest \cite{costi.94} 
(although, as discussed in Sec.~III~B, the use of a logarithmic 
discretization and a Gaussian broadening procedure can 
overestimate the widths of high energy peaks in spectral densities).

Returning to (\ref{eq:pw-approx}) we expect this
to be a valid approximation as long as orthogonality effects do not 
significantly affect the overlap matrix elements 
$\langle m_{F}'|m_{I,GS} \rangle_{N}$ appearing
in the above expression (and similar expressions for $C_{0r}(\omega)$).
When this occurs, as it will do for sufficiently low energies 
\cite{anderson.67}, 
an increasing number of final states $|m_{F}>_{N}$ will
be nearly orthogonal to the initial groundstate $|m_{I,GS}>_{N}$ and
it will be necessary to include higher energy states.
Information on higher energy states is contained in previous 
iterations within the NRG procedure. Thus if we evaluate the non--equilibrium
quantities approximately by using only one energy shell, then 
we will obtain results with a restricted range of validity.
The range of validity can be estimated: 
orthogonality effects between initial and
final states become important in the strong coupling regime, i.e. for
frequencies $\omega << T_{K}$, where $T_{K}$ is the low energy scale of
the AKM (or the renormalized tunneling amplitude, $\Delta_r$, 
in the language of the spin--boson model). However, the exponent 
with which the matrix elements
$\langle m_{F}'|m_{I,GS} \rangle_{N}$ vanish will also influence the
range of validity. We find, by keeping only one energy shell
for each frequency $\omega$, that 
$P(\omega)\sim C_{0r}(\omega)/\omega\sim |\omega|^{\frac{\alpha}{2}}, 
|\omega| << \Delta_r$ (noticeable in the low energy part of
our results in Fig.~1 and Fig.~2, presented in the next section). 
The vanishing of $P(\omega)$ with $\omega\rightarrow 0$ reflects the
orthogonality of the matrix elements in the expression for $P(\omega)$ and
we see that the exponent governing this is $\alpha/2$.
The exact behaviour of $P(\omega)$ and $C_{0r}(\omega)$ for 
$\omega\rightarrow 0$ is not rigorously known. In special cases,
such as at the Toulouse point, $\alpha=\frac{1}{2}$, it is known that
$P(\omega\rightarrow 0)\sim {[{\frac{C_{0r}(\omega)}
{\omega}}]}_{\omega\rightarrow 0} \sim \mbox{cnst}$, and
$P(t)\sim e^{- a t}, t\rightarrow \infty$. This type of exponential
relaxation is expected for other $\alpha$ in the range $0 < \alpha < 1$. 
In any case, we see that the approximate evaluation of the non--equilibrium
dynamics taking just one energy shell into account for each frequency is
expected to be accurate for weak dissipation (when the overlap exponent
is small) and for energies which are not too small relative 
to $\Delta_{r}$. In the context of macroscopic
quantum coherence experiments in SQUIDs, one is interested in the short
time dynamics for times up to approximately $1/\Delta_r$. The long time
behaviour $t>> 1/\Delta_r$ is of interest in other contexts (e.g. microscopic
two level systems), and for these
it will be necessary to carry out the full calculation, including higher
energy shells, for the non--equilibrium dynamics. We expect that such
a calculation will give results as accurate as those for
equilibrium quantities \cite{costi.96}.

\section{Results}
For weak dissipation, $\alpha<<1$, one expects damped oscillations of the
two level system at a frequency reduced relative to the bare tunneling
frequency, $\Delta$, due to the coupling to
the environment. From Fig.~1, which shows $P(\omega)$,
we see that this expected behaviour is reproduced by our method. The presence
of an inelastic peak in $P(\omega)$ at $\omega=\Delta_r^{*}$, indicates 
damped oscillations of frequency $\Delta_{r}^{*}$. The width, 
$\gamma_r^{*}$, of the inelastic peak gives the characteristic time 
$1/\gamma_{r}^{*}$ for the decay of the oscillations. For
$\alpha << 1$ we find a renormalized tunneling frequency 
$\Delta_{r}^{*}\approx\Delta_{r}<\Delta$ with
$\Delta_{r}=\Delta[\frac{\Delta}{\omega_{c}}]^{\frac{\alpha}{1-\alpha}}$,
and $\Delta_r$ is the relevant low energy scale for weak dissipation as 
obtained within the NIBA and several other approaches 
(\cite{note2,kehrein.96}).  
The larger renormalization of the tunneling amplitude with increasing
dissipation is clearly seen in Fig.~1. Qualitatively similar results are
found for the quantity $C_{0r}(\omega)/\omega$ (shown in Fig.~2).
For weak dissipation, the inelastic peak width 
$\gamma_{r}^{*}\sim \alpha \Delta_{r}^{*}$ vanishes linearly with 
$\alpha$ \cite{weiss.93}. 
\begin{figure}[h]
\vspace{0.2cm}
\centerline{\psfig{figure=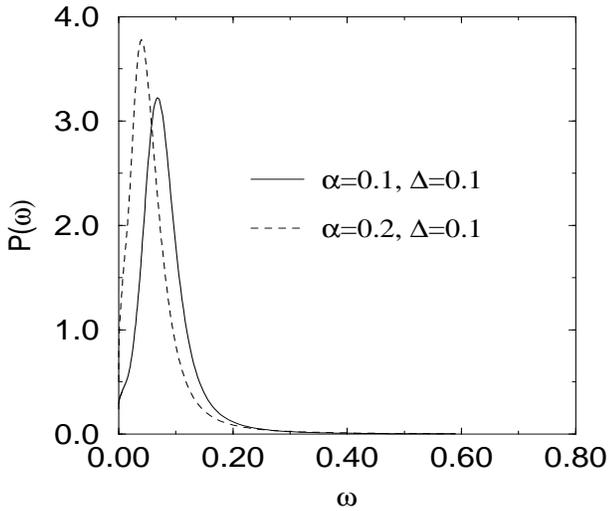,height=2.1in,width=2.3in}}
\vspace{1.2cm}
\caption{
$P(\omega)$ for $\Delta=0.1$ ($J_{\perp}=0.10397$)
$\alpha=0.1$ ($J_{\parallel}=4.698$), solid line, and $\alpha=0.2$ 
($J_{\parallel}=3.12759$), dashed line. 
Energies are in units of $D=\frac{\omega_{c}}{2}=1$. The relation
between the AKM parameters $J_{\perp,\parallel}$ and the spin--boson
model parameters $\alpha,\Delta$ is given in the text. Renormalizations
of these parameters due to the discretization of the AKM model have
been described in \protect\cite{costi.96}
}
\label{fig1}
\end{figure}
Consequently, for $\alpha << 1$, the use of
a logarithmic discretization does not provide the necessary resolution at
the peak position which is needed for resolving the intrinsic peak 
width, i.e. any broadening of the discrete spectra
will overestimate the inelastic peak width. This problem may be 
overcome by averaging 
over discretizations using the procedure in \cite{yoshida.90}. 
The standard procedure used here gives correctly the positions 
and weights of the delta functions in the discrete spectra 
(this has been shown in detail for all $\alpha$ in the 
range $0< \alpha < 1$ for the case of equilibrium dynamical quantities 
\cite{costi.96}). The decay of the weights of the delta functions with
increasing energy and hence the frequency dependence of $P(\omega)$ at
energies $\Delta_{r}<\omega < \omega_c$ is also correctly captured by
our procedure. We find that 
$P(\omega)\sim C_{0r}(\omega)/\omega \sim \omega^{-{(3-2\alpha)}}$ for
$\Delta << \omega_c$, $\Delta_r \ll \omega \ll \omega_c$ and
for $0<\alpha< 1$ (the error in the exponent is typically less than 0.1\%).
Thus, the short time behaviour of $P(t)$ is identical to the 
NIBA result $P(t)=1-c t^{2(1-\alpha)}$ for 
${\omega_c}^{-1}\ll t \ll {\Delta_r}^{-1}$. This gives independent 
confirmation that the NIBA is correct for $P(t)$ at short times.

Hence for weak dissipation we recover the known picture 
\cite{leggett.87,weiss.93} of damped oscillations at reduced
tunneling frequency $\Delta_{r}$. In particular, the short--time
dynamics is non--trivial in the sense that $P(t)$ and correlation
functions depend on $\alpha$ dependent exponents. 
On increasing the dissipation strength, the inelastic
peak in $P(\omega)$ narrows and the incoherent contribution 
($P(\omega=0)$) becomes larger.
For sufficiently strong dissipation, we expect the incoherent part
of $P(\omega)$ to dominate and lead to incoherent dynamics of the 
two level system. Although the question of the crossover
from coherent to incoherent dynamics in equilibrium quantities has been
investigated with high accuracy using the NRG \cite{costi.96}, 
the same question is 
technically more difficult for non--equilibrium quantities such as
$P(\omega)$. In this paper we have calculated the non--equilibrium 
dynamics only approximately, and the approximation used, which we 
described in detail in Sec.~III~C, has a range of validity which 
restricts us to weak dissipation and energies which are comparable to
or higher than the low energy scale of the model. To address the
question of a crossover from coherent to incoherent dynamics we
need to evaluate $P(\omega)$ taking into account several energy shells
for each frequency $\omega$, as detailed in Sec.~III~C.
\vspace{0.9cm}
\begin{figure}[h]
\centerline{\psfig{figure=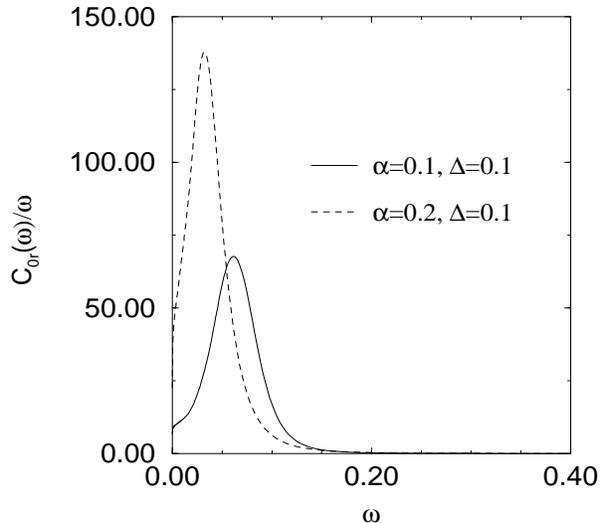,height=2.0in,width=2.0in}}
\vspace{1.0cm}
\caption{
$C_{0r}(\omega)$ for $\Delta=0.1$ ($J_{\perp}=0.10397$). (a)
$\alpha=0.1$ ($J_{\parallel}=4.698$), solid line, and (b) $\alpha=0.2$ 
($J_{\parallel}=3.12759$), dashed line. 
Energies are in units of $D=\frac{\omega_{c}}{2}=1$.
}
\label{fig2}
\end{figure}

\section{Conclusions}
To summarize, we have presented a numerical renormalization group 
approach to the calculation of non--equilibrium Green functions 
in correlated impurity systems. The approach uses an analogy to 
the calculation of response functions in the x--ray problem where 
the disturbance is sudden. The method was illustrated by
calculating the non--equilibrium dynamics of the Ohmic two--state system.
An approximate evaluation of the non--equilibrium quantities, 
taking only one energy shell into account for each frequency range,
gave accurate results within the expected range of validity of the 
approximation: specifically the approximate evaluation of $P(t)$ gave 
$P(t)=1-c t^{2(1-\alpha)}$ for 
$\omega_{c}^{-1} \ll t \ll \Delta_r^{-1}$ 
for all dissipation strengths provided $\Delta << \omega_c$. This is in
agreement with the NIBA prediction, which is known to be accurate at
short times. Including additional energy shells in the evaluation of 
non-equilibrium quantities should give essentially exact results. This is
left for future work as it requires a modified NRG procedure (outlined in
Sec.~III~C). The method is non--perturbative and
can be used to study the effects of local interactions on the non--equilibrium
transport through small devices such as quantum dots and tunnel junctions.
An important aspect of the method is that the disturbance (an electric
or magnetic field) can be arbitrarily large, so
it may be used to study the non--linear regime in the I--V 
characteristics of small devices, such as quantum dots.
\acknowledgments
We acknowledge useful discussions with P. W\"{o}lfle, 
Ph. Nozi\`eres and T. Martin. This work was supported by 
E.U. grant ERBCHRX CT93-0115 and the Institut Laue--Langevin.


\begin{references}
\bibitem[\ddagger]{tac-email}
e-mail: {\em tac@tkm.physik.uni-karlsruhe.de}
\bibitem{qds.95} {\em Quantum Dynamics of Submicron Structures}, edited by
H. A. Cerdeira, B. Kramer and G. Sch\"{o}n, NATO ASI Series, Vol.E291, 
(Kluwer Academic Publishers, Dordrecht, 1995). 
\bibitem{meso.91}{\em Mesoscopic Phenomena in Solids}, edited by 
B. Altshuler, P. A. Lee and R. Webb (North Holland, Amsterdam, 1991).
\bibitem{eaves.94} A. K. Geim et al., {\em Phys. Rev. Lett.}{\bf 72}, 
2061 (1994)
\bibitem{girvin.90} S. M. Girvin, L. I. Glazman, M. Jonson, D. R. Penn and
M. D. Stiles, {\em Phys. Rev. Lett.} {\bf 64}, 3183 (1990); M. H. Devoret, 
D. Esteve, H. Grabert, G.-L. Ingold, H. Pothier and C. Urbina, 
{Phys. Rev. Lett.} {\bf 64}, 1824 (1990).
\bibitem{matveev.92} K. A. Matveev and A. I. Larkin, {\em Phys. Rev.} 
{\bf B46}, 15337 (1992).
\bibitem{ng.88} T. K. Ng and P. A. Lee, {\em Phys. Rev. Lett.} {\bf 61}, 
1768 (1988).
\bibitem{glazmann.88} L. I. Glazmann and M. E. Raikh, 
{\em JETP Lett.} {\bf 47}, 452 (1988).
\bibitem{meir.91+93} Y. Meir, N. Wingreen and P. A. Lee,  
{\em Phys. Rev. Lett.} {\bf 66}, 3048 (1991); Y. Meir, 
N. Wingreen and P. A. Lee, {\em Phys. Rev. Lett.} {\bf 70}, 2601 (1993).
\bibitem{wingreen.94} N. Wingreen and Y. Meir,  {\em Phys. Rev.}
{\bf B49}, 11040 (1994).
\bibitem{hershfield.91+92} S. Hershfield, J. H. Davies and J. W. Wilkins,
{\em Phys. Rev. Lett.} {\bf 67}, 3720 (1991); 
S. Hershfield, J. H. Davies and J. W. Wilkins,
{\em Phys. Rev.} {\bf B46}, 7046 (1992).
\bibitem{schoeller.96} J. K\"{o}nig, H. Schoeller, and G. Sch\"{o}n 
{\em Phys. Rev. Lett.} {\bf 76}, 1715 (1996). 
\bibitem{keldysh.65} L. V. Keldysh, Sov. Phys. -- JETP {\bf 20}, 1018 (1965).
\bibitem{kadanoff.62} L. P. Kadanoff and G. Baym, {\em Quantum Statistical
Mechanics} (Benjamin, New York, 1962).
\bibitem{wilson.75} K.~G.~Wilson, Rev. Mod. Phys. {\bf 47}, 773 (1975).
\bibitem{kww.80} H. B. Krishnamurthy, J. W. Wilkins \&
K. G. Wilson, {\em Phys. Rev. {\bf B21}, 1044 (1980)}.
\bibitem{nd.69} P. Nozi\`eres and C. T. De Domenicis, 
{\em Phys. Rev. } {\bf 178}, 1097 (1969).
\bibitem{caldeira+leggett} A. O. Caldeira and A. J. Leggett, 
{\em Ann. Phys. (N.Y.)} {\bf 149}, 374 (1984); {\bf 153}, 445(E), (1984).
\bibitem{chakravarty.83} 
S. Chakravarty and S. Kivelson, 
{\em Phys. Rev. Lett.}{\bf 50}, 1811 (1983);
S. Chakravarty and A. J. Leggett, 
{\em Phys. Rev. Lett.}{\bf 52}, 5 (1984);
A. Garg, 
{\em Phys. Rev. }{\bf B 32}, 4746 (1985).
\bibitem{leggett.87} A. J. Leggett, S. Chakravarty, A. T. Dorsey,
M. P. A. Fisher, A. Garg and W. Zwerger, {\em Rev. Mod. Phys.} 
{\bf 59},1 (1987); {\bf 67}, 725 (1995).
\bibitem{weiss.93} U. Weiss, Series in 
Modern Condensed Matter Physics, Vol. 2, World Scientific, Singapore (1993).
\bibitem{toulouse.69} 
G. Toulouse, {\em C.R.Acad.Sci. Paris} 
{\bf 268}, 1200 (1969).
\bibitem{vigmann.78} P. B. Vigmann and A. M. Finkel'ste\u{i}n,
{\em Sov. Phys. JETP} {\bf 48},102 (1978); 
P. Schlottmann, {\em Phys. Rev.} {\bf B25}, 4805 (1982).
\bibitem{guinea.85b} F. Guinea, V. Hakim and A. Muramatsu, 
{\em Phys. Rev.} {\bf B32} 4410, (1985).
\bibitem{costi.96} T. A. Costi and C. Kieffer, {\em Phys. Rev. Lett.} 
{\bf 76}, 1683 (1996).
\bibitem{kotliar.96} G. Kotliar and Q. Si, {\em Phys. Rev.} {\bf B53} (1996).
\bibitem{anderson.70} P. W. Anderson, G. Yuval and D. R. Hamann,
{\em Phys. Rev.} {\bf B1} 4464, (1970).
\bibitem{tsvellick.83} A. M. Tsvellick and P. B. Wiegmann, {\em Adv. Phys.}
{\bf 32}, 453 (1983).
\bibitem{lesage.96} F. Lesage, H. Saleur and S. Skorik, {\em Phys. Rev. Lett.}
{\bf 76}, 3388 (1996).
\bibitem{costi.94} T. A. Costi, A. C. Hewson and V. Zlati\'{c} {\em J.
Phys.: Cond. Matt.} {\bf 6}, 2519 (1994); T. A. Costi, A. C. Hewson, 
{\em J. Phys.: Cond. Matt.} {\bf 5}, 361 (1993); 
{\em Phil. Mag.} B{\bf 65}, 1165 (1992). 
\bibitem{jones.87} This symmetry is a consequence of the local exchange
interactions and the particle--hole symmetry of the conduction band. It was
first used in NRG calculations in the context of the two--impurity Kondo
problem, B. Jones, C. M. Varma and J. W. Wilkins, {\em Phys. Rev. Lett.} 
{\bf 61}, 125 (1988). 
\bibitem{note1} This is due, first, to the absence of the above symmetries 
which would substantially increase the size of the matrices that need to
be diagonalized, and, second, in the corresponding linear chain model for
a bosonic heat bath each orbital $b_{n}$ (corresponding to $f_{n\mu}$ in 
the fermionic case) can be occupied with arbitrary number of bosons. This
would require an additional approximation to be made, restricting the
number of such bosons in each orbital $b_{n}$ to a finite number.
\bibitem{anderson.67} P. W. Anderson, {\em Phys. Rev. Lett.}{\bf 18},
1049 (1967).
\bibitem{yoshida.90} M. Yoshida, M. A. Whitaker and L. N. Oliveira, 
{\em Phys. Rev.} {\bf B 41}, 9403 (1990); W. C. Oliveira and L. N. Oliveira,
{\em Phys. Rev.} {\bf B 49}, 11986 (1994).
\bibitem{note2}
This energy scale also appears as a scaling invariant of the 
Anderson--Yuval scaling equations for the AKM in the limit 
$J_{\parallel}\rightarrow\infty$, which corresponds to 
$\alpha\rightarrow 0$ in the spin--boson model \cite{leggett.87,anderson.70}.
\bibitem{kehrein.96} S. K. Kehrein, A. Mielke and P. Neu, 
{\em Z. Phys.} {\bf B99}, 269 (1996); S. K. Kehrein and A. Mielke, 
{\em Preprint cond--mat/9607160}, Ann. Physik (Leipzig, 1996).
\end{references}
\end{document}